\documentclass[final]{aipproc}

\layoutstyle{8x11double}

\begin{document}

\title{GRB 060218 and the binaries as progenitors of GRB-SN systems}

\classification{98.70.Rz}
\keywords{GRBs and Binaries}

\author{Maria Giovanna Dainotti}{
  address={Dipartimento di Fisica, Universit\`a di Roma ``La Sapienza'', Roma, I-00185, Italy}
  ,altaddress={ICRANet and ICRA, Piazzale della Repubblica 10, Pescara, I-65122, Italy}
}

\author{Maria Grazia Bernardini}{
  address={Dipartimento di Fisica, Universit\`a di Roma ``La Sapienza'', Roma, I-00185, Italy}
  ,altaddress={ICRANet and ICRA, Piazzale della Repubblica 10, Pescara, I-65122, Italy}
}

\author{Carlo Luciano Bianco}{
  address={Dipartimento di Fisica, Universit\`a di Roma ``La Sapienza'', Roma, I-00185, Italy}
  ,altaddress={ICRANet and ICRA, Piazzale della Repubblica 10, Pescara, I-65122, Italy}
}
\author{Letizia Caito}{
address={Dipartimento di Fisica, Universit\`a di Roma ``La Sapienza'', Roma, I-00185, Italy}
  ,altaddress={ICRANet and ICRA, Piazzale della Repubblica 10, Pescara, I-65122, Italy}
}

\author{Roberto Guida}{
address={Dipartimento di Fisica, Universit\`a di Roma ``La Sapienza'', Roma, I-00185, Italy}
  ,altaddress={ICRANet and ICRA, Piazzale della Repubblica 10, Pescara, I-65122, Italy}
}

\author{Remo Ruffini}{
address={Dipartimento di Fisica, Universit\`a di Roma ``La Sapienza'', Roma, I-00185, Italy}
  ,altaddress={ICRANet and ICRA, Piazzale della Repubblica 10, Pescara, I-65122, Italy}
}

\begin{abstract}
We study the Gamma-Ray Burst (GRB) 060218: a particularly close source at $z=0.033$ with an extremely long duration, namely $T_{90}\sim 2000$ s, related to SN 2006aj. This source appears to be a very soft burst, with a peak in the spectrum at $4.9$ keV, therefore interpreted as an X-Ray Flash (XRF). It fullfills the Amati relation.
I present the fitting procedure, which is time consuming. In order to show its sensitivity I also present two examples of fits with the same value of $B$ and different value of $E_{e^\pm}^{tot}$. 
We fit the X- and $\gamma$-ray observations by \emph{Swift} of GRB 060218 in the $0.1$--$150$ keV energy band during the entire time of observations from $0$ all the way to $10^6$ s within a unified theoretical model.
The free parameters of our theory are only three, namely the total energy $E_{e\pm}^{tot}$ of the $e^\pm$ plasma, its baryon loading $B \equiv M_Bc^2/E_{e\pm}^{tot}$, as well as the CircumBurst Medium (CBM) distribution. 
We justify the extremely long duration of this GRB by a total energy $E_{e\pm}^{tot} = 2.32\times 10^{50}$ erg, a very high value of the baryon loading $B=1.0\times 10^{-2}$ and the effective CircumBurst Medium (CBM) density which shows a radial dependence $n_{cbm} \propto r^{-\alpha}$ with $1.0 \leq \alpha \leq 1.7$ and monotonically decreases from $1$ to $10^{-6}$ particles/cm$^3$. We recall that this value of the $B$ parameter is the highest among the sources we have analyzed and it is very close to its absolute upper limit expected.
 
By our fit we show that there is no basic differences between XRFs and more general GRBs. They all originate from the collapse process to a black hole and their difference is due to the variability of the three basic parameters within the range of full applicability of the theory. 
We also think that the smallest possible black hole, formed by the gravitational collapse of a neutron star in a binary system, is consistent with the especially low energetics of the class of GRBs associated with SNe Ib/c.
\end{abstract}

\maketitle

\section{Introduction}
GRB 060218, discovered by the \emph{Swift} satellite with cosmological redshift $z=0.033$ \cite{Mi06}, is one of the best examples of very long duration GRBs associated with core collapse Supernovae \cite{caa06}. We present a detailed fit of the X- and $\gamma$-ray luminosity in the entire time and energy band of observation, as well as details on the spectral distribution during the prompt emission phase. The additional peculiarities of the source evidence the lowest value of the Circumburst Medium (CBM) number density as well as the highest value of the baryon loading parameter yet observed in a GRB source. In this sense this source explores the applicability of GRB models in a yet untested range of physical parameters.

We present the observational data of the source; then we show the theoretical fit of the light curves observed by BAT and XRT ($15$--$150$ keV and $0.1$--$10.0$ keV respectively and the latest \textit{Chandra} observations showing possible analogies to the class of faint GRBs at low value of cosmological redshift, see Fig. \ref{060218chandra}). We show in Fig. \ref{spettro_completo_integrato} the predicted instantaneous spectrum from $100$ s (i.e. during the so called ``prompt emission'') all the way up to about $10^3$ s (i.e. until the gamma peak ends).  

We emphasize the aspects which makes this source so special: i) the total energy $E_{e\pm}^{tot} = 2.32\times 10^{50}$ erg, ii) the baryon loading parameter $B \equiv M_Bc^2/E_{e\pm}^{tot}= 1.0 \times 10^{-2}$ and iii) The effective CircumBurst Medium (CBM) density shows a radial dependence $n_{cbm}\propto r^{-\alpha}$ with $1.0 \leq \alpha \leq 1.7$ and monotonically decreases from $1$ to $10^{-6}$ particles/cm$^3$. We are going to show in the following how the occurrence of these three factors, within a consistent theoretical framework \cite{rlet1,rlet2,rubr,rubr2,EQTS_ApJL,EQTS_ApJL2,PowerLaws}, naturally leads to the explanation of the long duration ($T_{90}\sim 2000$ s) of this GRB.

In our approach we assume that all GRBs, short or long, originate from the gravitational collapse to a black hole \cite{rlet2}. We have only two free parameters describing the source, namely the total energy $E_{e\pm}^{tot}$ of the $e^\pm$ plasma and its baryon loading $B$ \cite{rswx00}. They characterize the optically thick adiabatic acceleration phase of the GRB, which lasts until the transparency condition is reached. After this acceleration phase, it starts the afterglow emission, due to the collision between the accelerated baryonic matter and the CBM. The CBM is described by two additional parameters: the effective particle number density $ \left\langle n_{CBM} \right\rangle  $ and the ratio between the effective emitting area and the total area of the pulse, ${\cal R} \equiv A_{eff} / A_{vis}$ \cite{spectr1}, which both takes into account the CBM filamentary structure \cite{fil}.
We emphasize the belonging to the Amati relation. 

\section{the observational properties of the source}

GRB 060218 has been triggered and located by the BAT instrument \cite{ca06} on board of the {\em Swift} satellite on 18 February 2006. It has a very long duration with $T_{90}\sim (2100 \pm 100)$s. The XRT \cite{ca06} began observations $\sim 153$ s after the BAT trigger{Ka06} and continued to detect the source for $\sim 12.3$ days \cite{Sa06}. The source is characterized by a flat gamma ray light curve and a soft spectrum \cite{Ba06}. It has an X-ray light curve with a long, slow rise and gradual decline and it is considered an X-ray flash since its peak energy occurs at $E_p=4.9^{+0.4}_{-0.3}$ keV \cite{caa06}. The burst fluence in the $15$--$150$ keV band is $(6.8\pm 0.4)\times 10^{-6}$ erg/cm$^2$ \cite{Sa06}. The spectroscopic redshift has been found to be $z=0.033$ \cite{Mi06}. At this redshift the isotropic equivalent energy is $E_{iso}=(1.9\pm 0.1)\times 10^{49}$ erg \cite{Sa06}. This faint, low redshift GRB could be a good candidate to be associated with a Supernova. In fact it has been found an underlying Type Ic Supernova: SN2006aj \cite{Pi06}. This Supernova shows observational features very similar to the other ones associated with GRBs. In particular it has a very large expansion velocity of $v\sim 0.1c$ \cite{Pi06,fa06,So06a}.

\section{The fit of the observed data}
\begin{figure}
  \includegraphics[height=.3\textheight]{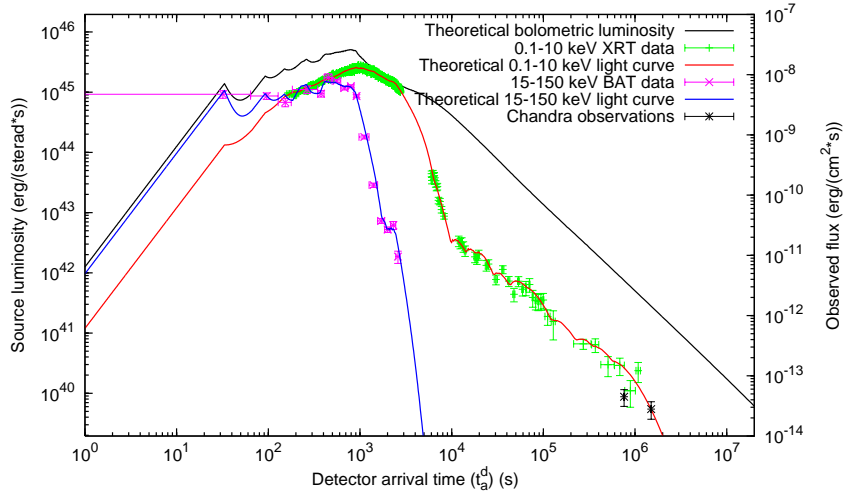}
  \caption{GRB 060218 complete light curves: our theoretical fit (blue line) of the $15$--$150$ keV BAT observations (pink points), our theoretical fit (red line) of the $0.1$--$10$ keV XRT observations (green points) and the $0.1$--$10$ keV \textit{Chandra} observations (black points) are represented together with our theoretically computed bolometric luminosity (black line) (data from \cite{caa06,So06b}).}
  \label{060218chandra}
\end{figure}

In this section we present the fit of our fireshell model to the observed data (see Fig. \ref{060218chandra}). The fit leads to a total energy of the $e^\pm$ plasma $E_{e^\pm}^{tot}= 2.32\times 10^{50}$ erg, with an initial temperature $T = 1.86$ MeV and a total number of pairs $N_{e^\pm} = 1.79\times 10^{55}$. The second parameter of the theory, $B = 1.0 \times 10^{-2}$, is close to the limit for the stability of the adiabatic optically thick acceleration phase of the fireshell (for further details see \cite{rswx00}). The Lorentz gamma factor obtained solving the fireshell equations of motion \cite{EQTS_ApJL2,PowerLaws} is $\gamma_\circ=99.2$ at the beginning of the afterglow phase at a distance from the progenitor $r_\circ=7.82\times 10^{12}$ cm. 

In Fig. \ref{060218chandra} we show the afterglow light curves fitting the prompt emission both in the BAT ($15$--$150$ keV) and in the XRT ($0.3$--$10$ keV) energy ranges, {\bf as expected in our ``canonical GRB'' scenario (see Dainotti et al., in preparation).} Initially the two luminosities are comparable to each other, but for a detector arrival time $t_a^d > 1000$ s the XRT curves becomes dominant. The displacement between the peaks of these two light curves leads to a theoretically estimated spectral lag greater than $500$ s in perfect agreement with the observations \cite{la06}. 

We recall that at $t_a^d \sim 10^4$ s there is a sudden enhancement in the radio luminosity and there is an optical luminosity dominated by the SN2006aj emission {\bf \cite{caa06,So06b}.} Although our analysis addresses only the BAT and XRT observations, for $r > 10^{18}$ cm corresponding to $t_a^d > 10^4$ s the fit of the XRT data implies two new features: \textbf{1)} a sudden increase of the ${\cal R}$ factor from ${\cal R} = 1.0\times 10^{-11}$ to ${\cal R} = 1.6\times 10^{-6}$, corresponding to a significantly more homogeneous effective CBM distribution; \textbf{2)} an XRT luminosity much smaller than the bolometric one (see Fig. \ref{060218chandra}). Therefore, we identify two different regimes in the afterglow, one for $t_a^d < 10^4$ s and the other for $t_a^d > 10^4$ s. Nevertheless, there is a unifying feature: the determined effective CBM density decreases with the distance $r$ monotonically and continuously through both these two regimes from $n_{cbm} = 1$ particle/cm$^3$ at $r = r_\circ$ to $n_{cbm} = 10^{-6}$ particle/cm$^3$ at $r = 6.0 \times 10^{18}$ cm: $n_{cbm} \propto r^{-\alpha}$, with $1.0 \leq \alpha \leq 1.7$.

\section{The procedure of the fit\label{procedure} }

\begin{figure}
  \includegraphics[height=.3\textheight]{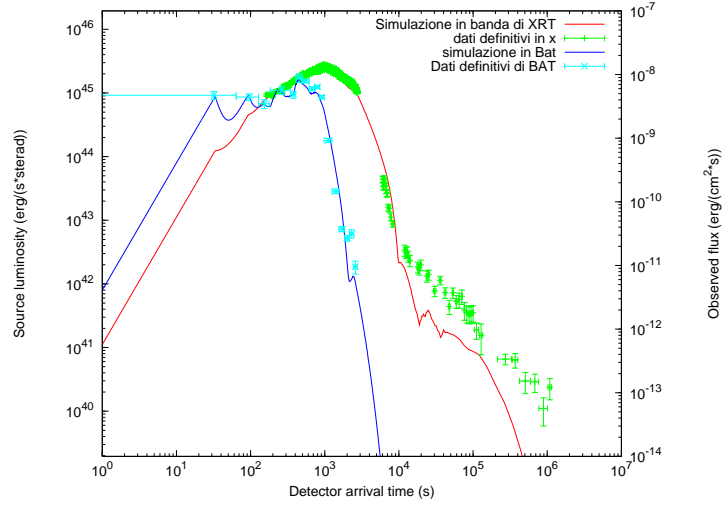}
  \caption{GRB060218 light curves with $E_{e^\pm}^{tot}= 1.36\times 10^{50}$ erg: our theoretical fit (blue line) of the $15$--$150$ keV BAT observations (pink points), our theoretical fit (red line) of the $0.3$--$10$ keV XRT observations (green points) and the $0.3$--$10$ keV \textit{Chandra} observations (black points) are represented together with our theoretically computed bolometric luminosity (black line) (Data from: \cite{caa06,So06b}).}
  \label{060218bolometricasotto}
\end{figure}

\begin{figure}
  \includegraphics[height=.3\textheight]{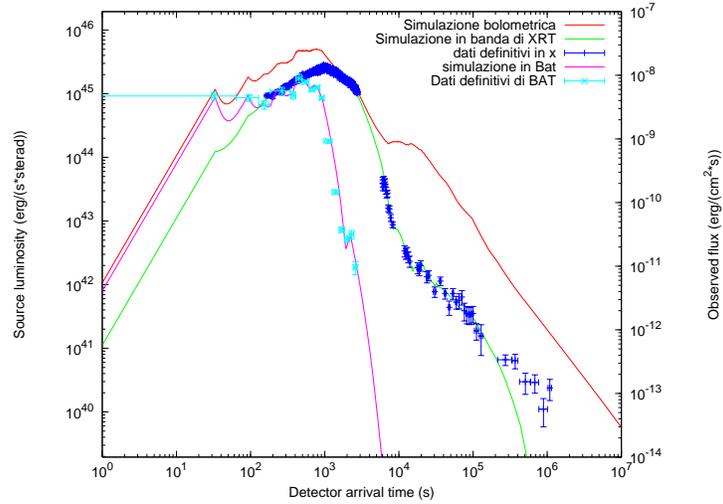}
  \caption{GRB060218 light curves with $E_{e^\pm}^{tot}= 1.61\times 10^{50}$ erg: our theoretical fit (blue line) of the $15$--$150$ keV BAT observations (pink points), our theoretical fit (red line) of the $0.3$--$10$ keV XRT observations (green points) and the $0.3$--$10$ keV \textit{Chandra} observations (black points) are represented together with our theoretically computed bolometric luminosity (black line). Data from: \cite{caa06,So06b}.}
  \label{060218senzaultimipunti}
\end{figure}

The arrival time of each photon at the detector depends on the entire previous history of the fireshell \cite{rlet1}. Moreover, all the observables depends on the EQTS \cite{EQTS_ApJL,EQTS_ApJL2} which, in turn, depend crucially on the equations of motion of the fireshell. The CBM engulfment has to be computed self-consistently through the entire dynamical evolution of the fireshell and not separately at each point. Any change in the CBM distribution strongly influences the entire dynamical evolution of the fireshell and, due to the EQTS structure, produces observable effects up to a much later time. For example if we change the density mask at a certain distance from the black hole we modify the shape of the lightcurve and consequently the evolution changes at larger radii corresponding to later times. 
Anyway the change of the density is not the only problem to face in the fitting of the source, in fact first of all we have to choose the energy in order to have Lorentz gamma factor sufficiently high to fit the entire GRB.  
In order to show the sensitivity of the fitting procedure I also present two examples of fits with the same value of $B$ and different value of $E_{e^\pm}^{tot}$.
The first example has an $E_{e^\pm}^{tot}$ = $1.36\times 10^{50}$ erg . This fit resulted unsuccessfully as we see from the Fig.\ref {060218bolometricasotto}, because the bolometric lightcurve is under the XRT peak of the afterglow. This means that the value of the energy chosen is too small to fit any data points after the peak of the afterglow. So we have to increase the value of the Energy to a have a better fit. In fact the parameters values have been found with various attempt in order to obtain the best fit. 
The second example is characterized by $E_{e^\pm}^{tot}= 1.61\times 10^{50}$ erg and the all the data are fitted except for the last point from $2.0\times 10^{2}$s to the end (see Fig. \ref{060218senzaultimipunti}). I attempt to fit these last points trying to diminuishes the $R$ values in order to enhance the energy emission, but again the low value of the Lorentz gamma factor, that in this case is $3$ prevent the fireshell to expand. So again in this case the value of the Energy chosen is too small, but it is better than the previous attempt. In this case we increased the energy value of the 24\%, but it is not enough so we decide to increase 16\%.
So the final fit is characterized by the $B=1.0\times10^{-2}$ and by the $E_{e^\pm}^{tot}= 2.32\times 10^{50}$ erg. With this value of the energy we are able to fit all the experimental points.

\section{Binaries as progenitors of GRB-SN systems}\label{binary}

The majority of the existing models in the literature appeal to a single astrophysical phenomenon to explain both the GRB and the SN \citep[``collapsar'', see e.g.][]{wb06}. On the contrary, a distinguishing feature of our theoretical approach is to differentiate between the SN and the GRB process. The GRB is assumed to occur during the formation process of a black hole. The SN is assumed to lead to the formation of a neutron star (NS) or to a complete disruptive explosion without remnants and, in no way, to the formation of a black hole. In the case of SN2006aj the formation of such a NS has been actually inferred by \citet{Mae06} because of the large amount of $^{58}$Ni ($0.05 M_\odot$). Moreover the significantly small initial mass of the SN progenitor star $M \approx 20 M_\odot$ is expected to form a NS rather than a black hole when its core collapses \citep{Mae06,Fe06,Maz06,No07}. In order to fulfill both the above requirement, we assume that the progenitor of the GRB and the SN consists of a binary system formed by a NS close to its critical mass collapsing to a black hole, and a companion star evolved out of the main sequence originating the SN. The temporal coincidence between the GRB and the SN phenomenon is explained in term of the concept of ``induced'' gravitational collapse \citep{rlet3,Mosca_Orale}. There is also the distinct possibility of observing the young born NS out of the SN \citep[see e.g.][and references therein]{Mosca_Orale}.

It has been often proposed that GRBs associated with SNe Ib/c, at smaller redshift $0.0085 < z < 0.168$ \citep[see e.g.][and references therein]{De06}, form a different class, less luminous and possibly much more numerous than the high luminosity GRBs at higher redshift \citep{Pi06,So04,Mae06,De06}. Therefore they have been proposed to originate from a separate class of progenitors \citep{Liang06,Cob06}. In our model this is explained by the nature of the progenitor system leading to the formation of the black hole with the smallest possible mass: the one formed by the collapse of a just overcritical NS \citep{RuffiniTF1,Mosca_Orale}.

The recent observation of GRB060614 at $z=0.125$ without an associated SN \citep{DV06,Mangano07} gives strong support to our scenario, alternative to the collapsar model. Also in this case the progenitor of the GRB appears to be a binary system composed of two NSs or a NS and a white dwarf (Caito et al., in preparation).

\section{Conclusions}\label{dicu}

GRB060218 presents a variety of peculiarities, including its extremely large $T_{90}$ and its classification as an XRF. 
The anomalously long $T_{90}$ led us to infer a monotonic decrease in the CBM effective density.
The spectrum from the prompt phase to the early part of the afterglow varies smoothly and continuously with characteristic hard to soft transition. 
Our scenario originates from the gravitational collapse to a black hole and is now confirmed over a $10^6$ range in energy. It is clear that, although the process of gravitational collapse is unique, there is a large variety of progenitors which may lead to the formation of black holes, each one with precise signatures in the energetics. The low energetics of the class of GRBs associated with SNe, and the necessity of the occurrence of the SN, naturally leads in our model to identify their progenitors with the formation of the smallest possible black hole originating from a NS overcoming his critical mass in a binary system. 
GRB060218 is the first GRB associated with SN with complete coverage of data from the onset all the way up to $\sim 10^6$ s. This fact offers an unprecedented opportunity to verify theoretical models on such a GRB class.

\end{document}